\documentclass[aps,preprintnumbers,superscriptaddress,nofootinbib,twocolumn]{revtex4}

\usepackage{amsthm}
\usepackage{amsmath}
\usepackage{amssymb}
\usepackage{amstext}
\usepackage{graphicx}
\usepackage{url}

\theoremstyle{plain}
\newtheorem{theorem}{Theorem}
\newtheorem{lemma}[theorem]{Lemma}
\newtheorem{corollary}[theorem]{Corollary}

\theoremstyle{definition}

\newcommand*{\comment}[1]{}
\newcommand*{\tr}{\mathrm{tr}}
\newcommand*{\ket}[1]{| #1 \rangle}
\newcommand*{\bra}[1]{\langle #1 |}
\newcommand*{\braket}[2]{\langle #1|#2\rangle}
\newcommand*{\ketbra}[2]{|#1\rangle\!\langle#2|}
\newcommand*{\spr}[2]{\langle #1 | #2 \rangle}
\newcommand*{\proj}[1]{\ket{#1}\bra{#1}}
\newcommand*{\id}{\openone}
\newcommand*{\ot}{\otimes}

\newcommand*{\genFid}{\bar{F}}

\newcommand*{\supp}{\mathrm{supp}}

\newcommand*{\eps}{\varepsilon}
\newcommand*{\cB}{\mathcal{B}}
\newcommand*{\cE}{\mathcal{E}}
\newcommand*{\cH}{\mathcal{H}}
\newcommand*{\cI}{\mathcal{I}}
\newcommand*{\cM}{\mathcal{M}}

\newcommand*{\cS}{\mathcal{S}}
\newcommand*{\cU}{\mathcal{U}}

\newcommand*{\obs}{R}
\newcommand*{\obst}{S}
\newcommand*{\cobs}{\mathcal{\obs}}
\newcommand*{\cobst}{\mathcal{\obst}}
\newcommand*{\Dobs}{D_\obs}
\newcommand*{\Dobst}{D_\obst}
\newcommand*{\sys}{A}
\newcommand*{\mem}{B}

\newcommand*{\HR}{H_{-\infty}}

\begin{document}

\title{The Uncertainty Principle in the Presence of Quantum
  Memory\footnote{The published version of this work can be found in
    {\it Nature Physics}~{\bf 6}, 659--662 (2010) at \url{http://www.nature.com/nphys/journal/vaop/ncurrent/abs/nphys1734.html}.}}

\author{Mario \surname{Berta}}
\affiliation{Institute for Theoretical Physics, ETH Zurich, 8093
Zurich, Switzerland.}
\affiliation{Faculty of Physics, Ludwig-Maximilians-Universit\"at M\"unchen, 80333 Munich,
Germany.}
\author{Matthias \surname{Christandl}}
\affiliation{Institute for Theoretical Physics, ETH Zurich, 8093
Zurich, Switzerland.}
\affiliation{Faculty of Physics, Ludwig-Maximilians-Universit\"at M\"unchen, 80333 Munich,
Germany.}
\author{Roger~\surname{Colbeck}}
\affiliation{Perimeter Institute for Theoretical Physics, 31 Caroline
Street North, Waterloo, ON N2L 2Y5, Canada.}
\affiliation{Institute for Theoretical Physics, ETH Zurich, 8093
Zurich, Switzerland.}
\affiliation{Institute of Theoretical Computer Science, ETH Zurich, 8092
Zurich, Switzerland.}
\author{Joseph M.\ \surname{Renes}}
\affiliation{Institute for Applied Physics, Technische Universit\"at Darmstadt,  64289 Darmstadt, Germany.}
\author{Renato \surname{Renner}}
\affiliation{Institute for Theoretical Physics, ETH Zurich, 8093
Zurich, Switzerland.}

\date{1st March 2011}

\maketitle

{\bf The uncertainty principle, originally formulated by
  Heisenberg~\cite{Heisenberg27}, dramatically illustrates the
  difference between classical and quantum mechanics.  The principle
  bounds the uncertainties about the outcomes of two incompatible
  measurements, such as position and momentum, on a particle.  It
  implies that one cannot predict the outcomes for both possible
  choices of measurement to arbitrary precision, even if information
  about the preparation of the particle is available in a classical
  memory.  However, if the particle is prepared entangled with a
  quantum memory, a device which is likely to soon be
  available~\cite{JulsgaardQuantumMemory}, it is possible to predict
  the outcomes for both measurement choices precisely.  In this work
  we strengthen the uncertainty principle to incorporate this case,
  providing a lower bound on the uncertainties which depends on the
  amount of entanglement between the particle and the quantum memory.
  We detail the application of our result to witnessing entanglement
  and to quantum key distribution.}\bigskip

Uncertainty relations constrain the potential knowledge one can have
about the physical properties of a system.  Although classical theory
does not limit the knowledge we can simultaneously have about
arbitrary properties of a particle, such a limit does exist in quantum
theory.  Even with a complete description of its state, it is
impossible to predict the outcomes of all possible measurements on the
particle.  This lack of knowledge, or uncertainty, was quantified by
Heisenberg~\cite{Heisenberg27} using the standard deviation (which we
denote by $\Delta\obs$ for an observable $\obs$).
If the measurement on a given particle is chosen from a set of two
possible observables, $\obs$ and $\obst$, the resulting bound on the
uncertainty can be expressed in terms of the
commutator~\cite{Robertson}:
\begin{equation*}
\Delta\obs\cdot\Delta\obst\geq\frac{1}{2}|\langle[\obs,\obst]\rangle|.
\end{equation*}
In an information-theoretic context, it is more natural to quantify
uncertainty in terms of entropy rather than the standard deviation.
Bia{\l}ynicki-Birula and Mycielski~\cite{BiaMyc75} derived entropic
uncertainty relations for position and momentum and
Deutsch~\cite{deutsch} later proved a relation that holds for any pair
of observables.  Subsequently, Kraus~\cite{kraus} conjectured an
improvement of Deutsch's result which was later proven by Maassen and
Uffink~\cite{MaaUff88}.  The improved relation is
\begin{equation}\label{eq:MaaUff}
H(\obs)+H(\obst)\geq \log_2 \frac{1}{c},
\end{equation}
where $H(\obs)$ denotes the Shannon entropy of the probability
distribution of the outcomes when $\obs$ is measured.  The term
$\frac{1}{c}$ quantifies the complementarity of the observables.  For
non-degenerate observables, $c := \max_{j, k}
|\spr{\psi_j}{\phi_k}|^2$ where $\ket{\psi_j}$ and $\ket{\phi_k}$ are
the eigenvectors of $\obs$ and $\obst$, respectively.

\begin{figure*}
\includegraphics[width=0.9\textwidth]{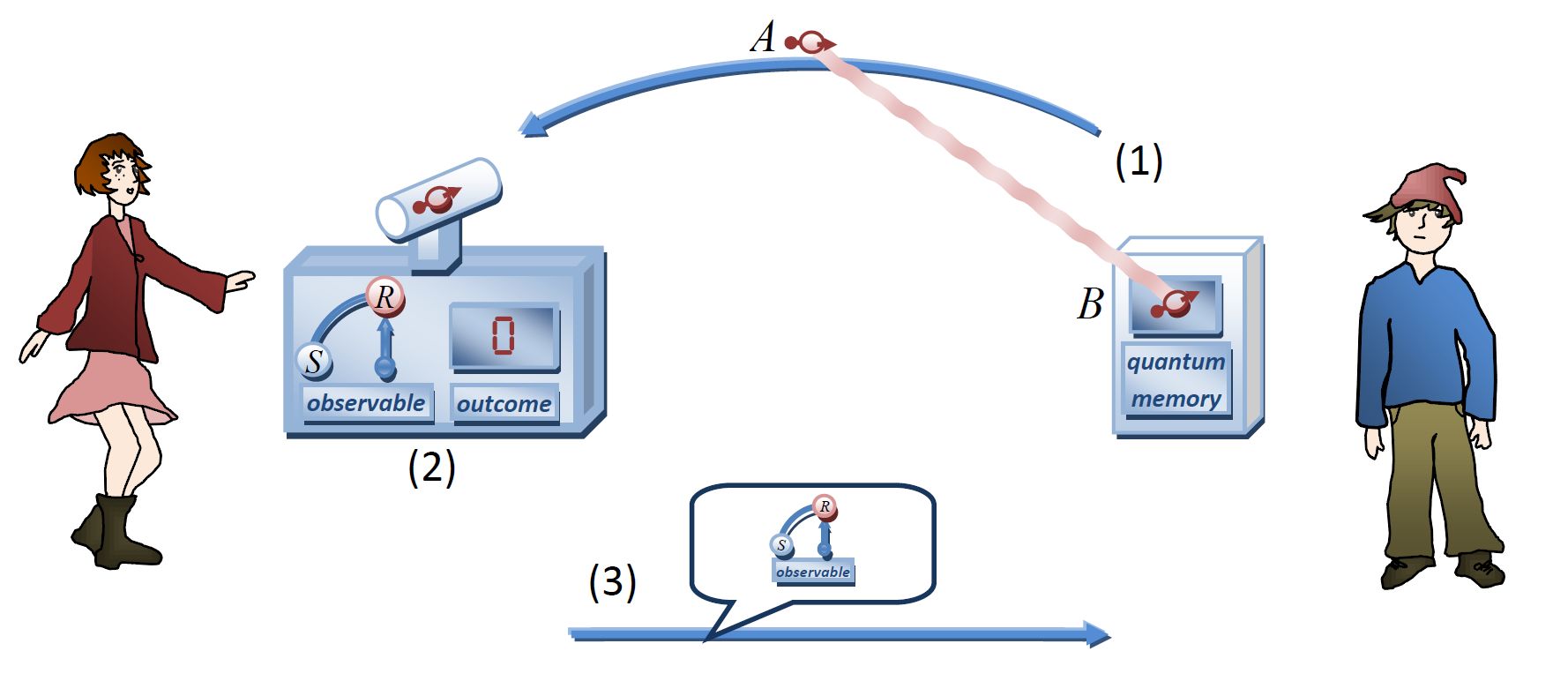}
\caption{Illustration of the uncertainty game. (1) Bob sends a
 particle to Alice, which may, in general, be entangled with his
 quantum memory.  (2) Alice measures either $\obs$ or $\obst$ and
 notes her outcome.  (3) Alice announces her measurement choice to
 Bob.  Our uncertainty relation provides a lower bound on Bob's
 resulting uncertainty about Alice's outcome.}
\label{fig:1}
\end{figure*}

One way to think about uncertainty relations is via the following game
(the uncertainty game) between two players, Alice and Bob.  Before the
game commences, Alice and Bob agree on two measurements, $\obs$ and
$\obst$.  The game proceeds as follows: Bob prepares a particle in a
quantum state of his choosing and sends it to Alice.  Alice then
performs one of the two measurements and announces her choice to Bob.
Bob's task is to minimize his uncertainty about Alice's measurement
outcome.  This is illustrated in Figure~\ref{fig:1}.

Equation~\eqref{eq:MaaUff} bounds Bob's uncertainty in the case that
he has no quantum memory|all information Bob holds about the particle
is classical, e.g., a description of its density matrix.  However,
with access to a quantum memory, Bob can beat this bound.  To do so,
he should maximally entangle his quantum memory with the particle he
sends to Alice.  Then, for any measurement she chooses, there is a
measurement on Bob's memory which gives the same outcome as Alice
obtains.  Hence, the uncertainties about both observables, $\obs$ and
$\obst$, vanish, which shows that if one tries to generalize
Equation~\eqref{eq:MaaUff} by replacing the measure of uncertainty
about $\obs$ and $\obst$ used there (the Shannon entropy) by the
entropy \emph{conditioned on the information in Bob's quantum memory},
the resulting relation no longer holds.

We proceed by stating our uncertainty relation, which applies in the
presence of a quantum memory. It provides a bound on the uncertainties
of the measurement outcomes which depends on the amount of
entanglement between the measured particle, $\sys$, and the quantum
memory, $\mem$.  Mathematically, it is the relation
\begin{equation}\label{eq:main}
H(\obs|\mem)+H(\obst|\mem)\geq\log_2\frac{1}{c}+H(\sys|\mem).
\end{equation}
The uncertainty about the outcome of measurement $\obs$ given
information stored in a quantum memory, $\mem$, is denoted by the
conditional von Neumann entropy, $H(\obs|\mem)$.
The additional term $H(\sys|\mem)$ appearing on the right hand side
quantifies the amount of entanglement between the particle and the
memory.  We sketch the proof of this relation in the Methods section
and defer the full proof to the Supplementary Information.

We continue by discussing some instructive examples:
\begin{enumerate}
\item If the particle, $\sys$, and memory, $\mem$, are maximally
 entangled, then $H(\sys|\mem)=-\log_2 d$, where $d$ is the dimension
 of the particle sent to Alice.  Since $\log_2\frac{1}{c}$ cannot
 exceed $\log_2 d$, the bound~\eqref{eq:main} reduces to
 $H(\obs|\mem)+H(\obst|\mem)\geq 0$, which is trivial, since the
 conditional entropy of a system after measurement given the quantum
 memory cannot be negative.  As discussed above, Bob can guess both
 $\obs$ and $\obst$ perfectly with such a strategy.

\item If $\sys$ and $\mem$ are not entangled (i.e., their state is a
 convex combination of product states) then $H(\sys|\mem)\geq
 0$. Since $H(\obs|\mem) \leq H(\obs)$ and $H(\obst|\mem) \leq
 H(\obst)$ for all states, we recover Maassen and Uffink's bound,
 Equation~\eqref{eq:MaaUff}.

\item In the absence of the quantum memory, $\mem$, we can reduce the
bound~\eqref{eq:main} to
$H(\obs)+H(\obst)\geq\log_2\frac{1}{c}+H(\sys)$.  If the state of
the particle, $\sys$, is pure, then $H(\sys)=0$ and we again recover
the bound of Maassen and Uffink, Equation~\eqref{eq:MaaUff}.
However, if the particle, $\sys$, is in a mixed state then $H(\sys)>0$
and the resulting bound is stronger than Equation~\eqref{eq:MaaUff}
even when there is no quantum memory.

\item In terms of new applications, the most interesting case is when
$\sys$ and $\mem$ are entangled, but not maximally so.  Since a
negative conditional entropy $H(\sys|\mem)$ is a signature of
entanglement~\cite{DevWin}, the uncertainty relation takes into
account the entanglement between the particle and the memory.  It is
therefore qualitatively different from existing classical bounds.

\end{enumerate}

Aside from its fundamental significance, our result has impact on the
development of future quantum technologies. In the following we will
explain how it can be applied to the task of witnessing entanglement
and to construct security proofs in quantum cryptography.

For the application to witnessing entanglement, consider a source
which emits a two-particle state $\rho_{\sys\mem}$.  Analogously to
the uncertainty game, we measure $\sys$ with one of two observables,
$\obs$ or $\obst$.  Furthermore, a second measurement (of $\obs'$ or
$\obst'$) should be applied to $\mem$ trying to reproduce the outcome
of the first. The probability with which the measurements on $\sys$
and $\mem$ disagree can be directly used to upper bound the entropies
$H(\obs|\mem)$ and $H(\obst|\mem)$.  For example, using Fano's
inequality, we obtain $H(\obs|\mem) \leq
h(p_{\obs})+p_{\obs}\log_2(d-1)$, where $p_{\obs}$ is the probability
that the outcomes of $\obs$ and $\obs'$ are not equal and $h$ is the
binary entropy function.  If this bound and the analogous bound for
$H(\obst|\mem)$ are sufficiently small, then our
result,~\eqref{eq:main}, implies that $H(\sys|\mem)$ must be negative,
and hence that $\rho_{\sys\mem}$ is entangled.

Note that this method of witnessing entanglement does not involve a
(usually experimentally challenging) estimation of the $D^2$ matrix
elements of $\rho_{\sys\mem}$, where $D$ is the dimension of
$\sys\mem$|it is sufficient to estimate the two probabilities
$p_{\obs}$ and $p_{\obst}$, which can be obtained by separate
measurements on each of the two particles.  Our method also differs
significantly from the standard approach which is based on collecting
measurement statistics to infer the expectation values of fixed
\emph{witness observables} on the joint system of both
particles~\cite{Horodeckis,Terhal,optwit,wit_review}.  We remark that
when using our procedure, the best choice of Alice's observables are
ones with high complementarity, $\frac{1}{c}$.

As a second application, we consider quantum key distribution.  In the
1970s and 80s, Wiesner~\cite{Wiesner70}, and Bennett and
Brassard~\cite{BB84} proposed new cryptographic protocols based on
quantum theory, most famously the BB84 quantum key distribution
protocol~\cite{BB84}.  Their intuition for security lay in the
uncertainty principle.  In spite of providing the initial intuition,
the majority of security proofs to date have not involved uncertainty
relations (see e.g.~\cite{DEJMPS, LoCha99, ShoPre00, christandl-2004,
 RenKoe05, RRthesis}), although~\cite{koashi-2005} provides a notable
exception.  The obstacle for the use of the uncertainty principle is
quickly identified: a full proof of security must take into account a
technologically unbounded eavesdropper, i.e.\ one who potentially has
access to a quantum memory.  In the following, we explain how to use
our main result,~\eqref{eq:main}, to overcome this obstacle and
derive a simple bound on the key rate.

Based on an idea by Ekert~\cite{Ekert91}, the security of quantum key
distribution protocols is usually analysed by assuming that the
eavesdropper creates a quantum state, $\rho_{ABE}$, and distributes the
$A$ and $B$ parts to the two users, Alice and Bob.  In practice, Alice
and Bob do not provide the eavesdropper with this luxury, but a
security proof that applies even in this case will certainly imply
security when Alice and Bob distribute the states themselves.  In
order to generate their key, Alice and Bob measure the states they
receive using measurements chosen at random, with Alice's possible
measurements denoted by $R$ and $S$ and Bob's by $R'$ and $S'$.  To
ensure that the same key is generated, they communicate their
measurement choices to one another.  In the worst case, this
communication is overheard in its entirety by the eavesdropper who is
trying to obtain the key.  Even so, Alice and Bob can generate a
secure key if their measurement outcomes are sufficiently well
correlated.

To show this, we use a result of Devetak and Winter~\cite{DevWin} who
proved that the amount of key Alice and Bob are able to extract per
state, $K$, is lower bounded by $H(R|E)-H(R|B)$.  In addition, we
reformulate our main result,~\eqref{eq:main}, as $H(R|E)+H(S|B)\geq
\log_2\frac{1}{c}$, a form previously conjectured by Boileau and
Renes~\cite{RenBoi09} (see the Supplementary Information).  Together
these imply $K\geq \log_2\frac{1}{c}-H(R|B)-H(S|B)$.  Furthermore,
using the fact that measurements cannot decrease entropy, we have
\begin{align*}
K\geq \log_2\frac{1}{c}-H(R|R')-H(S|S').
\end{align*}
This corresponds to a generalization of Shor and Preskill's famous
result~\cite{ShoPre00}, which is recovered in the case of conjugate
observables applied to qubits and assuming symmetry, i.e.\
$H(R|R')=H(S|S')$.  The argument given here applies only to collective
attacks but can be extended to arbitrary attacks using the
post-selection technique~\cite{Christandl2009}.

This security argument has the advantage that Alice and Bob only need
to upper bound the entropies $H(R|R')$ and $H(S|S')$.  Similarly to
the case of entanglement witnessing, these entropies can be directly
bounded by observable quantities, such as the frequency with which
Alice and Bob's outcomes agree.  No further information about the
state is required.  This improves the performance of practical quantum
key distribution schemes, where the amount of statistics needed to
estimate states is critical for security~\cite{RenSca08}.

The range of application of our result,~\eqref{eq:main}, is not
restricted to these two examples, but extends to other cryptographic
scenarios~\cite{fehr}, a quantum phenomenon known as \emph{locking of
 information}~\cite{locking} (in the way presented
in~\cite{ChrWin05}), and to \emph{decoupling theorems} which are
frequently used in coding arguments~\cite{RenBoi09}. 

Finally, we note that uncertainty may be quantified in terms of
alternative entropy measures.  In fact, our proof involves
\emph{smooth entropies}, which can be seen as generalizations of the von
Neumann entropy~\cite{RRthesis} (see the Methods and Supplementary
Information).  These generalizations have direct operational
interpretations~\cite{KRS} and are related to physical quantities,
such as thermodynamic entropy.  We therefore expect a formulation of
the uncertainty relation in terms of these generalized entropies to
have further use both in quantum information theory and beyond.

\mbox{}

\section*{Methods}
Here we outline the proof of the main result,~\eqref{eq:main}.  The
quantities appearing there are evaluated for a state
$\rho_{\sys\mem}$, where we use $H(\obs|\mem)$ to denote the
conditional von Neumann entropy of the state
$$\sum_j(\ketbra{\psi_j}{\psi_j}\ot\id)\rho_{AB}(\ketbra{\psi_j}{\psi_j}\ot\id),$$
and likewise for $H(\obst|\mem)$.

The proof is fully based on the \emph{smooth entropy calculus}
introduced in~\cite{RRthesis} and proceeds in three steps (we refer
the reader to the Supplementary Information for further details,
including precise definitions of the quantities used in this
section). In the first step, which we explain in more detail below, an
uncertainty relation is proven which is similar to~\eqref{eq:main} but
with the von Neumann entropy being replaced by the min- and
max-entropies, denoted $H_{\min}$ and $H_{\max}$ (we also use $\HR$ which plays a role similar to $H_{\max}$):
\begin{align} \label{eq:min}
H_{\min}(\obs|\mem) + \HR(\obst \mem) \geq \log_2\frac{1}{c} + H_{\min}(\sys\mem)\ .
\end{align}
The quantities $\HR$ and $H_{\min}$ only involve the extremal
eigenvalues of an operator, which makes them easier to deal with than
the von Neumann entropy which depends on all eigenvalues.  In the
second, technically most involved step of the proof, we extend the
relation to the \emph{smooth} min- and max-entropies, which are more
general and allow us to recover the relation for the von Neumann
entropy as a special case.

The $\eps$-smooth min- and max-entropies are formed by taking the
original entropies and extremizing them over a set of states
$\eps$-close to the original (where closeness is quantified in terms
of the maximum purified distance from the original).  In this step we
also convert $\HR$ to a smooth max-entropy and obtain the relation
\begin{align}\nonumber
H_{\min}^{5 \sqrt{\eps}}(\obs|\mem)+H_{\max}^{\eps}(\obst \mem) \geq\\ \log_2\frac{1}{c} + &H_{\min}^\eps(\sys\mem)- 2 \log_2 \frac{1}{\eps} \ ,\label{eq:smooth}
\end{align}
which holds for any $\eps>0$.

To complete the proof, we evaluate the inequality on the $n$-fold
tensor product of the state in question, i.e.\ on $\rho^{\otimes
n}$. We then use the asymptotic equipartition
theorem~\cite{RRthesis,ToCoRe08}, which tells us that the smooth min-
and max-entropies tend to the von Neumann entropy in the appropriate
limit, i.e.\
\begin{align*}
\lim_{\eps\rightarrow
0}\lim_{n\rightarrow\infty}\frac{1}{n}H_{\min/\max}^{\eps}(\sys^n|\mem^n)_{\rho^{\ot
 n}}=H(\sys|\mem)_{\rho}.
\end{align*}
Hence, on both sides of~\eqref{eq:smooth}, we divide by $n$ and take
the limit as in the previous equation to obtain
$$H(\obs|\mem)+H(\obst \mem)\geq\log_2\frac{1}{c}+H(\sys\mem),$$ from
which our main result, \eqref{eq:main}, follows by subtracting
$H(\mem)$ from both sides.

We now sketch the first step of the proof.  This develops an idea
from~\cite{ChrWin05, RenBoi09} where uncertainty relations which only
apply to the case of complementary observables (i.e.\ those related by
a Fourier transform) are derived.  These relations were originally
expressed in terms of von Neumann entropies rather than min- and
max-entropies.

We use two chain rules and strong subadditivity of the min-entropy, to
show that, for a system composed of subsystems $\sys'\mem'\sys\mem$
and for a state $\Omega$,
\begin{align}
H_{\min}(\sys' \mem' \sys \mem)_{\Omega}&-\HR(\sys' \sys \mem)_{\Omega} \label{eq:first}\\
\stackrel{\text{chain 1}}{\leq} &  H_{\min}(\mem'|\sys' \sys
\mem)_{\Omega|\Omega} \nonumber\\
\stackrel{\text{str.sub.}}{\leq}& H_{\min}(\mem'|\sys \mem)_{\Omega|\Omega} \nonumber\\
\stackrel{\text{chain 2}}{\leq} & H_{\min}(\mem' \sys|\mem)_{\Omega}-H_{\min}(\sys|\mem)_{\Omega}.\label{eq:combinedchain}
\end{align}

We now apply this relation to the state $\Omega_{\sys' \mem' \sys \mem}$ defined
as follows:
\begin{eqnarray*}
\Omega_{\sys' \mem' \sys \mem}   &:=& \frac{1}{d^2} \sum_{a, b} \proj{a}_{\sys'} \ot\proj{b}_{\mem'} \otimes\\
&&  (\Dobs^a \Dobst^b\ot\id)\rho_{\sys\mem}(\Dobst^{-b}\Dobs^{-a}\ot\id) \ ,
\end{eqnarray*}
where $\{\ket{a}\}_a$ and $\{\ket{b}\}_b$ are orthonormal bases on
$d$-dimensional Hilbert spaces $\cH_{\sys'}$ and $\cH_{\mem'}$
respectively, and $\Dobs$ and $\Dobst$ are the operators that dephase in the
respective eigenbases of $\obs$ and $\obst$. Hence, tracing out
$\sys'$ ($\mem'$) reduces the state to one where the system, $\sys$,
is measured in the eigenbasis of $\obs$ ($\obst$) and the outcome
forgotten.  We then use properties of the entropies to relate the
second term in~\eqref{eq:first} and the first term
in~\eqref{eq:combinedchain} to $\HR(\obst\mem)$ and
$H_{\min}(\obs|\mem)$, respectively, in spite of the fact that $\obs$
and $\obst$ neither commute nor anticommute|a property that makes it
difficult to complete the proof directly with the von Neumann entropy.
The first term of~\eqref{eq:first} is easily related to
$H_{\min}(\sys\mem)$.  Finally, tracing out both $\sys'$ and $\mem'$
reduces the state to one where the system, $\sys$, is measured first
with one observable and then with the other and the outcomes
forgotten.  Hence, the last term in~\eqref{eq:combinedchain} can be
related to the overlap of the eigenvectors of the two observables,
$c$.

Bringing everything together, we obtain the desired uncertainty
relation~\eqref{eq:min}.

\mbox{}

{\bf Acknowledgements:} We thank Robert K\"onig, Jonathan Oppenheim
and Marco Tomamichel for useful discussions and L\'idia del Rio for
the illustration (Figure~\ref{fig:1}).  MB and MC acknowledge support
from the German Science Foundation (DFG) and the Swiss National
Science Foundation. JMR acknowledges the support of CASED
(www.cased.de). RC and RR acknowledge support from the Swiss National
Science Foundation.




\begin{widetext}
\appendix
\setcounter{equation}{0}

\section*{Supplementary Information}
Here we present the full proof of our main result, the uncertainty
relation given in Equation~(2) of the main manuscript
(Theorem~\ref{thm:uncertaintyrelation} below).

In order to state our result precisely, we introduce a few
definitions. Consider two measurements described by orthonormal bases
$\{\ket{\psi_j}\}$ and $\{\ket{\phi_k}\}$ on a $d$-dimensional Hilbert
space $\cH_A$ (note that they are not necessarily complementary). The
measurement processes are then described by the completely positive
maps
\begin{align*}
 \cobs & : \rho \mapsto
 \sum_j\bra{\psi_j}\rho\ket{\psi_j}\ketbra{\psi_j}{\psi_j} \,\,\text{ and}\\
 \cobst & : \rho \mapsto \sum_k\bra{\phi_k}\rho\ket{\phi_k}\ketbra{\phi_k}{\phi_k}
\end{align*}
respectively. We denote the square of the overlap of these
measurements by $c$, i.e.\
\begin{equation}\label{eq:cdef}
 c := \max_{j, k} |\spr{\psi_j}{\phi_k}|^2.
\end{equation}
Furthermore, we assume that $\cH_B$ is an arbitrary finite-dimensional
Hilbert space.  The von Neumann entropy of $A$ given $B$ is denoted
$H(A|B)$ and is defined via $H(A|B):=H(AB)-H(B)$, where for a state
$\rho$ on $\cH_A$ we have $H(A):=-\text{tr}(\rho\log\rho)$.

The statement we prove is then
\begin{theorem}
\label{thm:uncertaintyrelation}
For any density operator $\rho_{AB}$ on
$\cH_A\ot\cH_B$,
\begin{equation}\label{eq:uncertaintyrelationthm}
H(\obs|\mem)+H(\obst|\mem)\geq\log_2\frac{1}{c}+H(\sys|\mem),
\end{equation}
where $H(\obs|B)$, $H(\obst|B)$, and $H(A|B)$ denote the conditional von
Neumann entropies of the states $(\cobs\ot\cI)(\rho_{AB})$,
$(\cobst\ot\cI)(\rho_{AB})$, and $\rho_{AB}$, respectively.
\end{theorem}

In the next section, we introduce the smooth min- and max- entropies
and give some properties that will be needed in the proof.

Before that, we show that the statement of our main theorem is
equivalent to a relation conjectured by Boileau and
Renes~\cite{RenBoi09s}.
\begin{corollary}
For any density operator $\rho_{ABE}$ on $\cH_A\ot\cH_B\ot\cH_E$,
\begin{equation}\label{eq:BR}
H(\obs|E) +  H(\obst|B) \geq \log_2\frac{1}{c}.
\end{equation}
\end{corollary}
\begin{proof}
To show that our result implies~\eqref{eq:BR}, we first
rewrite~\eqref{eq:uncertaintyrelationthm} as $H(\obs B) + H(\obst B)
\geq \log_2\frac{1}{c} + H(AB) + H(B)$.  In the case that $\rho_{A B
  E}$ is pure, we have $H(\obs B) = H(\obs E)$ and $H(AB)=H(E)$.  This
yields the expression $H(\obs E) + H(\obst B) \geq \log_2\frac{1}{c} +
H(E) + H(B)$, which is equivalent to~\eqref{eq:BR}. The result for
arbitrary states $\rho_{A B E}$ follows by the concavity of the
conditional entropy (see e.g.~\cite{NieChu}).

That~\eqref{eq:BR} implies~\eqref{eq:uncertaintyrelationthm} can be
seen by taking $\rho_{ABE}$ as the state which purifies $\rho_{AB}$
in~\eqref{eq:BR} and reversing the argument above.
\end{proof}

\subsection{(Smooth) min- and max-entropies|definitions}
\label{sec:defs}

As described above, we prove a generalized version
of~\eqref{eq:uncertaintyrelationthm}, which is formulated in terms of
smooth min- and max-entropies. This section contains the basic
definitions, while Section~\ref{app:entprops} summarizes the
properties of smooth entropies needed for this work. For a more
detailed discussion of the smooth entropy calculus, we refer
to~\cite{RRthesiss,KoReSc08,ToCoRe08s,ToCoRe09}.

We use $\cU_=(\cH):=\{\rho: \rho\geq 0, \tr\rho=1\}$ to denote the set
of normalized states on a finite-dimensional Hilbert space $\cH$ and
$\cU_\leq(\cH):=\{\rho: \rho\geq 0, \tr\rho\leq1\}$ to denote the set
of subnormalized states on $\cH$.  The definitions below apply to
subnormalized states.

The conditional min-entropy of $A$ given $B$ for a state
$\rho\in\cU_\leq(\cH_{AB})$ is defined as\footnote{In the case of
 finite dimensional Hilbert spaces (as in this work), the infima and
 suprema used in our definitions can be replaced by minima and
 maxima.}
\begin{align*}
 H_{\min}(A|B)_{\rho} := \sup_{\sigma} H_{\min}(A|B)_{\rho|\sigma}\ ,
\end{align*}
where the supremum is over all normalized density operators
$\sigma\in\cU_=(\cH_B)$ and where
$$H_{\min}(A|B)_{\rho|\sigma} := -\log_2 \inf \{\lambda: \, \rho_{AB} \leq \lambda \, \id_A \otimes \sigma_B \} \ .$$
In the special case where the $B$ system is trivial, we write
$H_{\min}(A)_{\rho}$ instead of $H_{\min}(A|B)_{\rho}$. It is easy to
see that $H_{\min}(A)_{\rho} = -\log_2 \| \rho_A \|_{\infty}$ and that
for $\rho\leq\tau$, $H_{\min}(A|B)_{\rho}\geq H_{\min}(A|B)_{\tau}$.

Furthermore, for $\rho\in\cU_\leq(\cH_A)$, we define
$$H_{\max}(A)_\rho := 2\log_2 \tr\sqrt{\rho}\ .$$  It follows
that for $\rho\leq\tau$, $H_{\max}(A)_{\rho}\leq H_{\max}(A)_{\tau}$
(since the square root is operator monotone).

In our proof, we also make use of an intermediate quantity, denoted
$\HR$. It is defined by
$$\HR(A)_\rho := -\log_2 \sup \{\lambda : \, \rho_A \geq \lambda \,
\Pi_{\supp(\rho_A)} \} \, ,$$ where $\Pi_{\supp(\rho_A)}$ denotes the
projector onto the support of $\rho_A$.  In other words,
$\HR(A)_{\rho}$ is equal to the negative logarithm of the smallest
non-zero eigenvalue of $\rho_A$.  This quantity will not appear in our
final statements but will instead be replaced by a smooth version of
$H_{\max}$ (see below and Section~\ref{app:entprops}).

The \emph{smooth} min- and max-entropies are defined by extremizing
the non-smooth entropies over a set of nearby states, where our notion
of nearby is expressed in terms of the \emph{purified distance}.  It
is defined as (see~\cite{ToCoRe09})
\begin{align} \label{eq:distdef}
 P(\rho,\sigma) := \sqrt{1-\genFid(\rho,\sigma)^2} \ ,
\end{align}
where $\genFid(\cdot\,,\cdot)$ denotes the generalized fidelity (which
equals the standard fidelity if at least one of the states is normalized),
\begin{align} \label{eq:genfidelity}
 \genFid(\rho,\sigma) := \bigl\| \sqrt{\rho\oplus (1-\tr \rho)} \sqrt{\sigma
   \oplus (1-\tr \sigma)} \bigr\|_1 \ .
\end{align}
(Note that we use $F(\rho,\sigma):=\|\sqrt{\rho}\sqrt{\sigma}\|_1$ to
denote the standard fidelity.)

The purified distance is a distance measure; in particular, it
satisfies the triangle inequality $P(\rho,\sigma)\leq
P(\rho,\tau)+P(\tau,\sigma)$. As its name indicates, $P(\rho, \sigma)$
corresponds to the minimum trace distance\footnote{The trace distance
 between two states $\tau$ and $\kappa$ is defined by $\frac{1}{2} \|
 \tau - \kappa \|_1$ where $\| \Gamma \|_1 = \tr \sqrt{\Gamma
   \Gamma^{\dagger}}$.} between purifications of $\rho$ and $\sigma$.
Further properties are stated in Section~\ref{app:purdist}.

We use the purified distance to specify a ball of subnormalized
density operators around $\rho$:
\begin{equation*}
 \cB^{\eps}(\rho):=\{\rho':\rho'\in\cU_\leq(\cH), P(\rho,\rho')\leq\eps \} \ .
\end{equation*}
Then, for any $\eps \geq 0$, the \emph{$\eps$-smooth min- and
 max-entropies} are defined by
\begin{eqnarray*}
 H_{\min}^\eps(A|B)_\rho &:=& \sup_{\rho' \in \cB^\eps(\rho)}
 H_{\min}(A|B)_{\rho'} \\
 H_{\max}^\eps(A)_\rho &:=& \inf_{\rho' \in \cB^\eps(\rho)}
 H_{\max}(A)_{\rho'} \ .
\end{eqnarray*}
In the following, we will sometimes omit the subscript $\rho$ when it
is obvious from context which state is implied.

\subsection{Overview of the proof} \label{sec:smoothrelation}

The proof of our main result, Theorem~\ref{thm:uncertaintyrelation},
is divided into two main parts, each individually proven in the next
sections.

In the first part, given in Section~\ref{sec:proof1}, we prove the
following uncertainty relation, which is similar to the main result
but formulated in terms of the quantum entropies $H_{\min}$ and $\HR$.
\begin{theorem} \label{thm:uncertaintyrelationnonsmooth} For any
 $\rho_{AB}\in\cU_\leq(\cH_{AB})$ we have
\begin{align*}
 H_{\min}(\obs|B)_{(\cobs \otimes \cI)(\rho)} + \HR(\obst B)_{(\cobst \otimes \cI)(\rho)} \geq \log_2\frac{1}{c} + H_{\min}(AB)_{\rho} \ .
\end{align*}
\end{theorem}

The second part of the proof involves \emph{smoothing} the above
relation and yields the following theorem (see
Section~\ref{sec:sm})\footnote{We note that a related relation
  follows from the work of Maassen and Uffink \cite{MaaUff88s} who
  derived a relation involving R\'enyi entropies (the order $\alpha$
  R\'enyi entropy \cite{Renyi} is denoted $H_{\alpha}$) and the
  overlap $c$ (defined in~\eqref{eq:cdef}).  They showed that
  $H_\alpha(\obs)_\rho+H_\beta(\obst)_\rho\geq\log_2\frac{1}{c}$,
  where $\frac{1}{\alpha}+\frac{1}{\beta}=2$.  The case
  $\alpha\rightarrow\infty$, $\beta\rightarrow\frac{1}{2}$ yields
  $H_{\min}(\obs)_\rho+H_{\max}(\obst)_\rho\geq\log_2\frac{1}{c}$.}.
\begin{theorem}\label{thm:uncertaintyrelationsmooth}
 For any $\rho\in\cU_=(\cH_{AB})$ and $\eps>0$,
\begin{align*}
H_{\min}^{5 \sqrt{\eps}}(\obs|B)_{(\cobs \otimes \cI)(\rho)}+H_{\max}^{\eps}(\obst B)_{(\cobst \otimes \cI)(\rho)} \geq \log_2\frac{1}{c} + H_{\min}^\eps(AB)_{\rho}- 2 \log_2 \frac{1}{\eps} \ .
\end{align*}
\end{theorem}

From Theorem~\ref{thm:uncertaintyrelationsmooth}, the von Neumann
version of the uncertainty relation
(Theorem~\ref{thm:uncertaintyrelation}) can be obtained as an
asymptotic special case for i.i.d.\ states.  More precisely, for any
$\sigma\in\cU_=(\cH_{AB})$ and for any $n \in \mathbb{N}$, we evaluate
the inequality for $\rho = \sigma^{\otimes n}$ where $\cobs\ot\cI$ and
$\cobst\ot\cI$ are replaced by $(\cobs\ot\cI)^{\ot n}$ and
$(\cobst\ot\cI)^{\ot n}$, respectively.  Note that the
 corresponding overlap is then given by
\begin{align*}
 c^{(n)} = \max_{j_1\ldots j_n, k_1\ldots k_n}
 |\spr{\psi_{j_1}}{\phi_{k_1}}\ldots\spr{\psi_{j_n}}{\phi_{k_n}}|^2=\max_{j,
   k} |\spr{\psi_j^{\ot n}}{\phi_k^{\ot n}}|^2 = c^n \ .
\end{align*}
The assertion of the theorem can thus be rewritten as
\begin{align*}
 \frac{1}{n} H_{\min}^{5 \sqrt{\eps}}(\obs^n|B^n)_{((\cobs \otimes \cI)(\sigma))^{\otimes n}} +\frac{1}{n}
 H_{\max}^{\eps}(\obst^n B^n)_{((\cobst \otimes \cI)(\sigma))^{\otimes n}}  \geq \log_2\frac{1}{c} + \frac{1}{n} H_{\min}^\eps(A^n B^n)_{\sigma^{\otimes n}} - \frac{2}{n} \log_2\frac{1}{\eps} \ .
\end{align*}
Taking the limit $n \to \infty$ and then $\eps \to 0$ and using the
asymptotic equipartition property (Lemma~\ref{lem:aep}), we obtain
$H(\obs|B)+H(\obst B)\geq \log_2\frac{1}{c}+H(AB)$, from which
Theorem~\ref{thm:uncertaintyrelation} follows by subtracting $H(B)$
from both sides.

\subsection{Proof of Theorem~\ref{thm:uncertaintyrelationnonsmooth}}
\label{sec:proof1}

In this section we prove a version of Theorem~\ref{thm:uncertaintyrelation},
formulated in terms of the quantum entropies $H_{\min}$ and
$\HR$.

 We introduce $\Dobs=\sum_j e^{\frac{2 \pi i
     j}{d}}\ketbra{\psi_j}{\psi_j}$ and $\Dobst=\sum_k e^{\frac{2 \pi i
     k}{d}}\ketbra{\phi_k}{\phi_k}$ ($\Dobs$ and $\Dobst$ are $d$-dimensional
 generalizations of Pauli operators).  The maps $\cobs$ and $\cobst$
 describing the two measurements can then be rewritten as
\begin{align*}
 \cobs & : \rho \mapsto \frac{1}{d}\sum_{a=0}^{d-1} \Dobs^a \rho \Dobs^{-a} \\
 \cobst & : \rho \mapsto \frac{1}{d}\sum_{b=0}^{d-1} \Dobst^b \rho \Dobst^{-b} \ .
\end{align*}
We use the two chain rules proved in Section~\ref{app:entprops}
(Lemmas~\ref{lem:chain1} and~\ref{lem:chain2}), together with the
strong subadditivity of the min-entropy (Lemma~\ref{lem:subadd}), to
obtain, for an arbitrary density operator $\Omega_{A' B' A B}$,
\begin{eqnarray}\nonumber
 H_{\min}(A' B' A B)_{\Omega}-\HR(A' A B)_{\Omega}&\leq&H_{\min}(B'|A' A B)_{\Omega|\Omega}\\
 \nonumber&\leq&H_{\min}(B'|A B)_{\Omega|\Omega}\\
 &\leq&H_{\min}(B' A|B)_{\Omega}-H_{\min}(A|B)_{\Omega}.\label{eq:combinedchains}
\end{eqnarray}

We now apply this relation to the state $\Omega_{A' B' A B}$ defined
as follows\footnote{The idea behind the use of this state first
 appeared in \cite{ChrWin05s}.}:
\begin{align*}
 \Omega_{A' B' A B} := \frac{1}{d^2} \sum_{a, b} \proj{a}_{A'} \ot\proj{b}_{B'} \otimes(\Dobs^a \Dobst^b\ot\id)\rho_{AB}(\Dobst^{-b}\Dobs^{-a}\ot\id) \ ,
\end{align*}
where $\{\ket{a}_{A'}\}_a$ and $\{\ket{b}_{B'}\}_b$ are orthonormal
bases on $d$-dimensional Hilbert spaces $\cH_{A'}$ and $\cH_{B'}$.

This state satisfies the following relations:
\begin{eqnarray}
\label{eq:omega1}H_{\min}(A' B' A B)_{\Omega}&=&2\log_2 d+H_{\min}(AB)_{\rho}\\
\label{eq:omega2}\HR(A' A B)_{\Omega}&=&\log_2 d+\HR(\obst B)_{(\cobst \otimes \cI)(\rho)}\\
\label{eq:omega3}H_{\min}(B' A|B)_{\Omega}&\leq&\log_2 d+H_{\min}(\obs|B)_{(\cobs \otimes \cI)(\rho)}\\
\label{eq:omega4}H_{\min}(A|B)_{\Omega}& \geq & \log_2\frac{1}{c} .
\end{eqnarray}
Using these in~\eqref{eq:combinedchains} establishes
Theorem~\ref{thm:uncertaintyrelationnonsmooth}.  We proceed by showing
\eqref{eq:omega1}--\eqref{eq:omega4}.

Relation~\eqref{eq:omega1} follows because $\Omega_{A' B' A B}$ is
unitarily related to $\frac{1}{d^2} \sum_{a, b} \proj{a}_{A'} \otimes
\proj{b}_{B'} \ot\rho_{AB}$, and the fact that the unconditional
min-entropy is invariant under unitary operations.

To see~\eqref{eq:omega2}, note that $\Omega_{A' A B}$ is unitarily
related to $\frac{1}{d^2} \sum_a
\proj{a}_{A'}\ot\sum_b(\obst^b\ot\id)\rho_{AB}(\obst^{-b}\ot\id)$ and that
$\frac{1}{d}\sum_b(\obst^b\ot\id)\rho_{AB}(\obst^{-b}\ot\id)=(\cobst\ot\cI)(\rho_{AB})$.

To show inequality~\eqref{eq:omega3}, note that
$$\Omega_{B' A B}=\frac{1}{d^2} \sum_b
\proj{b}_{B'}\ot\sum_a(\Dobs^a \Dobst^b\ot\id)\rho_{AB}(\Dobst^{-b}\Dobs^{-a}\ot\id).$$
To evaluate the min-entropy, define $\lambda$ such that $H_{\min}(B'
A|B)_\Omega=-\log_2\lambda$.  It follows that there exists a
(normalized) density operator $\sigma_B$ such that
\begin{eqnarray*}
\lambda \, \id_{B' A}\ot\sigma_B&\geq&\frac{1}{d^2} \sum_b
\proj{b}_{B'}\ot\sum_a(\Dobs^a \Dobst^b\ot\id)\rho_{AB}(\Dobst^{-b}\Dobs^{-a}\ot\id).
\end{eqnarray*}
Thus, for all $b$,
\begin{eqnarray*}
\lambda \, \id_{A}\ot\sigma_B&\geq&\frac{1}{d^2}\sum_a(\Dobs^a \Dobst^b\ot\id)\rho_{AB}(\Dobst^{-b}\Dobs^{-a}\ot\id),
\end{eqnarray*}
and in particular, for $b=0$, we have
\begin{eqnarray*}
\lambda \, \id_{A}\ot\sigma_B&\geq&\frac{1}{d^2}\sum_a(\Dobs^a\ot\id)\rho_{AB}(\Dobs^{-a}\ot\id)\\
&=&\frac{1}{d}(\cobs\ot\cI)(\rho_{AB}).
\end{eqnarray*}
We conclude that $2^{-H_{\min}(\obs|B)_{(\cobs \otimes
   \cI)(\rho)}}\leq\lambda d$, from which~\eqref{eq:omega3} follows.

To show~\eqref{eq:omega4}, we observe that
\begin{align*}
 \Omega_{A B}
=
 \frac{1}{d^2}\sum_{ab}(\Dobs^a \Dobst^b\ot\id)\rho_{AB}(\Dobst^{-b}\Dobs^{-a}\ot\id)
=
 ((\cobs \circ \cobst) \otimes \cI)(\rho_{A B}) \ .
\end{align*}
Then,
\begin{eqnarray*}
 ((\cobs \circ \cobst) \otimes \cI)(\rho_{AB})&=&(\cobs\ot\cI)\left(\sum_k\ketbra{\phi_k}{\phi_k}\ot\tr_A((\ketbra{\phi_k}{\phi_k}\ot\id)\rho_{AB})\right)\\
 &=&\sum_{j k}|\braket{\phi_k}{\psi_j}|^2\,\ketbra{\psi_j}{\psi_j}\ot\tr_A((\ketbra{\phi_k}{\phi_k}\ot\id)\rho_{AB})\\
&\leq&\max_{l m}\left(|\braket{\phi_l}{\psi_m}|^2\right)\,\sum_{j k}\ketbra{\psi_j}{\psi_j}\ot\tr_A((\ketbra{\phi_k}{\phi_k}\ot\id)\rho_{AB})\\
&=&\max_{l m}\left(|\braket{\phi_l}{\psi_m}|^2\right)\,\id_A\ot\sum_k\tr_A((\ketbra{\phi_k}{\phi_k}\ot\id)\rho_{AB})\\
&=&\max_{l m}\left(|\braket{\phi_l}{\psi_m}|^2\right)\,\id_A\ot\rho_B.
\end{eqnarray*}
It follows that $2^{-H_{\min}(A|B)_{((\cobs \circ \cobst) \otimes
   \cI)(\rho)}}\leq\max_{l m}|\braket{\phi_l}{\psi_m}|^2=c$, which
concludes the proof.
\qed

\subsection{Proof of Theorem~\ref{thm:uncertaintyrelationsmooth}}
\label{sec:sm}

The uncertainty relation proved in the previous section
(Theorem~\ref{thm:uncertaintyrelationnonsmooth}) is formulated in
terms of the entropies $H_{\min}$ and $\HR$. In this section, we
transform these quantities into the smooth entropies $H_{\min}^\eps$
and $H_{\max}^\eps$, respectively, for some $\eps>0$.  This will
complete the proof of Theorem~\ref{thm:uncertaintyrelationsmooth}.

Let $\sigma_{AB}\in\cU_\leq(\cH_{AB})$. Lemma~\ref{lem:smoothHR} applied to
$\sigma_{\obst B}:= (\cobst\ot\cI)(\sigma_{AB})$ implies that there exists a
nonnegative operator $\Pi\leq\id$ such that $\tr((\id-\Pi^2)\sigma_{\obst
 B})\leq 3\eps$ and
\begin{align} \label{eq:HmaxHRappl}
 H_{\max}^\eps(\obst B)_{(\cobst \otimes \cI)(\sigma)}\geq \HR(\obst B)_{\Pi (\cobst \otimes \cI)(\sigma) \Pi} - 2
 \log_2\frac{1}{\eps} \ .
\end{align}
We can assume without loss of generality that $\Pi$ commutes with the
action of $\cobst\ot\cI$ because it can be chosen to be diagonal in any
eigenbasis of $\sigma_{\obst B}$. Hence, $\Pi (\cobst \otimes \cI)(\sigma_{A
 B}) \Pi = (\cobst \otimes \cI)(\Pi \sigma_{A B} \Pi)$, and
\begin{align} \label{eq:tracePcommute}
 \tr((\id-\Pi^2)\sigma_{A B}) = \tr((\cobst \otimes \cI)((\id - \Pi^2) \sigma_{A B})) = \tr((\id-\Pi^2)\sigma_{\obst B}) \leq 3 \eps \ .
\end{align}
Applying Theorem~\ref{thm:uncertaintyrelationnonsmooth} to the
operator $\Pi\sigma_{AB}\Pi$ yields
\begin{align} \label{eq:uncertverone}
 H_{\min}(\obs|B)_{(\cobs \otimes \cI)(\Pi\sigma\Pi)} + H_{R}(\obst
 B)_{(\cobst \otimes \cI)(\Pi \sigma \Pi)} \geq \log_2\frac{1}{c} +
 H_{\min}(AB)_{\Pi \sigma \Pi} \ .
\end{align}
Note that $\Pi\sigma\Pi\leq\sigma$ and so
\begin{align} \label{eq:Hminproject}
 H_{\min}(AB)_{\Pi\sigma\Pi} \geq H_{\min}(AB)_{\sigma}.
\end{align}
Using~\eqref{eq:HmaxHRappl} and \eqref{eq:Hminproject} to bound the terms in~\eqref{eq:uncertverone}, we find
\begin{align}\label{eq:rhoexpr}
 H_{\min}(\obs|B)_{(\cobs \otimes \cI)(\Pi\sigma\Pi)} + H_{\max}^\eps(\obst
 B)_{(\cobst \otimes \cI)(\sigma)} \geq \log_2\frac{1}{c} + H_{\min}(AB)_{\sigma}
 - 2 \log_2 \frac{1}{\eps} \ .
\end{align}
Now we apply Lemma~\ref{lem:smoothHmin} to $\rho_{AB}$. Hence there
exists a nonnegative operator $\bar{\Pi} \leq \id$ which is diagonal
in an eigenbasis of $\rho_{A B}$ such that
\begin{align} \label{eq:distancetwo}
 \tr((\id- \bar{\Pi}^2) \rho_{A B}) \leq 2 \eps
\end{align}
and $H_{\min}(AB)_{\bar{\Pi} \rho \bar{\Pi}}\geq
H_{\min}^\eps(AB)_{\rho}$.  Evaluating~\eqref{eq:rhoexpr} for
$\sigma_{A B} := \bar{\Pi} \rho_{AB} \bar{\Pi}$ thus gives
\begin{align} \label{eq:uncertverfour}
 H_{\min}(\obs|B)_{(\cobs \otimes \cI)(\Pi \bar{\Pi} \rho \bar{\Pi} \Pi)}
 + H_{\max}^\eps(\obst B)_{(\cobst \otimes \cI)(\bar{\Pi} \rho \bar{\Pi})}
 \geq \log_2\frac{1}{c} + H_{\min}^\eps(AB)_{\rho} - 2 \log_2 \frac{1}{\eps} \ ,
\end{align}
where $\Pi$ is diagonal in any eigenbasis of $(\cobst \otimes
\cI)(\bar{\Pi} \rho_{A B} \bar{\Pi})$ and satisfies
\begin{align} \label{eq:distanceone} \tr((\id-\Pi^2) \bar{\Pi} \rho_{A
   B} \bar{\Pi}) \leq 3 \eps \ .
\end{align}
Since $\rho_{A B} \geq \bar{\Pi} \rho_{A B} \bar{\Pi}$, we can apply
Lemma~\ref{lem:Hzerosmoothsubstate} to $(\cobst \otimes \cI)(\rho_{A B})$
and $(\cobst \otimes \cI)(\bar{\Pi} \rho_{A B} \bar{\Pi})$, which gives
\begin{align} \label{eq:Hmaxepsboundu}
 H_{\max}^{\eps}(\obst B)_{(\cobst \otimes \cI)(\rho)} \geq H_{\max}^{\eps}(\obst B)_{(\cobst \otimes \cI)(\bar{\Pi} \rho \bar{\Pi})} \ .
\end{align}
The relation~\eqref{eq:uncertverfour} then reduces to
\begin{align} \label{eq:presmooth}
 H_{\min}(\obs|B)_{(\cobs \otimes \cI)(\Pi \bar{\Pi}\rho\bar{\Pi} \Pi)} +
 H_{\max}^{\eps}(\obst B)_{(\cobst \otimes \cI)(\rho)} \geq \log_2\frac{1}{c}  +
 H_{\min}^\eps(AB)_{\rho} - 2 \log_2 \frac{1}{\eps} \ .
\end{align}
Finally, we apply Lemma~\ref{lem:distprojection}
to~\eqref{eq:distancetwo} and~\eqref{eq:distanceone}, which gives
\begin{align*}
 P(\rho_{A B}, \bar{\Pi} \rho_{A B} \bar{\Pi}) & \leq \sqrt{4 \eps} \\
 P( \bar{\Pi} \rho_{A B} \bar{\Pi}, \Pi \bar{\Pi} \rho_{A B} \bar{\Pi} \Pi) & \leq \sqrt{6 \eps} \ .
\end{align*}
Hence, by the triangle inequality
\begin{align*}
 P(\rho_{AB}, \Pi \bar{\Pi} \rho_{AB} \bar{\Pi} \Pi) \leq
 (\sqrt{4} + \sqrt{6}) \sqrt{\eps} < 5 \sqrt{\eps} \ .
\end{align*}
Consequently, $(\cobs \otimes \cI)(\Pi \bar{\Pi} \rho_{AB} \bar{\Pi}
\Pi)$ has at most distance $5 \sqrt{\eps}$ from $(\cobs \otimes
\cI)(\rho_{A B})$. This implies
\begin{align*}
 H_{\min}^{5 \sqrt{\eps}}(\obs|B)_{(\cobs \otimes \cI)(\rho)}\geq
 H_{\min}(\obs|B)_{(\cobs \otimes \cI)(\Pi \bar{\Pi}\rho \bar{\Pi} \Pi)} \ .
\end{align*}
Inserting this in~\eqref{eq:presmooth} gives
$$H_{\min}^{5 \sqrt{\eps}}(\obs|B)_{(\cobs \otimes \cI)(\rho)}+H_{\max}^{\eps}(\obst B)_{(\cobst \otimes \cI)(\rho)} \geq \log_2\frac{1}{c} + H_{\min}^\eps(AB)_{\rho}- 2 \log_2 \frac{1}{\eps} \ ,
$$
which completes the proof of
Theorem~\ref{thm:uncertaintyrelationsmooth}.
\qed

\subsection{Technical properties}
\subsubsection{Properties of the purified distance} \label{app:purdist}

The purified distance between $\rho$ and $\sigma$ corresponds to the
minimum trace distance between purifications of $\rho$ and $\sigma$,
respectively~\cite{ToCoRe09}. Because the trace distance can only
decrease under the action of a partial trace (see, e.g.,
\cite{NieChu}), we obtain the following bound.

\begin{lemma} \label{lem:Ptracebound}
 For any $\rho\in\cU_\leq(\cH)$ and $\sigma\in\cU_\leq(\cH)$,
 \begin{align*}
   \|\rho-\sigma\|_1 \leq 2 P(\rho, \sigma) .
 \end{align*}
\end{lemma}

The following lemma states that the purified distance is
non-increasing under certain mappings.

\begin{lemma} \label{lem:Pnonincreasing} For any
 $\rho\in\cU_\leq(\cH)$ and $\sigma\in\cU_\leq(\cH)$, and for any
 nonnegative operator $\Pi \leq \id$,
\begin{align}
 P(\Pi \rho \Pi, \Pi \sigma \Pi) \leq P(\rho, \sigma).
\end{align}
\end{lemma}

\begin{proof}
 We use the fact that the purified distance is non-increasing under
 any trace-preserving completely positive map (TPCPM)~\cite{ToCoRe09}
 and consider the TPCPM $$\cE:
 \rho\mapsto\Pi\rho\Pi\oplus\tr(\sqrt{\id-\Pi^2}\rho\sqrt{\id-\Pi^2}).$$
 We have $P(\rho,\sigma)\geq P(\cE(\rho),\cE(\sigma))$, which implies
 $\genFid(\rho,\sigma)\leq\genFid(\cE(\rho),\cE(\sigma))$.  Then,
 \begin{eqnarray}\nonumber
   \genFid(\rho,\sigma)&\leq&\genFid(\cE(\rho),\cE(\sigma))\\\nonumber
   &=&F(\Pi\rho\Pi,\Pi\sigma\Pi)+\sqrt{(\tr\rho-\tr(\Pi^2\rho))(\tr\sigma-\tr(\Pi^2\sigma))}+\sqrt{(1-\tr\rho)(1-\tr\sigma)}\\\nonumber
   &\leq&F(\Pi\rho\Pi,\Pi\sigma\Pi)+\sqrt{(1-\tr(\Pi^2\rho))(1-\tr(\Pi^2\sigma))}\\\nonumber
   &=&\genFid(\Pi\rho\Pi,\Pi\sigma\Pi),
 \end{eqnarray}
which is equivalent to the statement of the Lemma.

The second inequality is the relation
\begin{eqnarray*}
\sqrt{(\tr\rho-\tr(\Pi^2\rho))(\tr\sigma-\tr(\Pi^2\sigma))}+\sqrt{(1-\tr\rho)(1-\tr\sigma)}&\leq&\sqrt{(1-\tr(\Pi^2\rho))(1-\tr(\Pi^2\sigma))},
\end{eqnarray*}
which we proceed to show.  For brevity, we write
$\tr\rho-\tr(\Pi^2\rho)=r$, $\tr\sigma-\tr(\Pi^2\sigma)=s$,
$1-\tr\rho=t$ and $1-\tr\sigma=u$.  We hence seek to
show $$\sqrt{rs}+\sqrt{tu}\leq\sqrt{(r+t)(s+u)}.$$ For $r$, $s$, $t$
and $u$ nonnegative, we have
\begin{eqnarray*}
\sqrt{rs}+\sqrt{tu}\leq\sqrt{(r+t)(s+u)}&\Leftrightarrow&rs+2\sqrt{rstu}+tu\leq(r+t)(s+u)\\
&\Leftrightarrow&4rstu\leq(ru+st)^2\\
&\Leftrightarrow&0\leq(ru-st)^2.
\end{eqnarray*}
\end{proof}

Furthermore, the purified distance between a state $\rho$ and its
image $\Pi\rho\Pi$ is upper bounded as follows.

\begin{lemma} \label{lem:distprojection} For any
 $\rho\in\cU_\leq(\cH)$, and for any nonnegative operator,
 $\Pi\leq\id$,
$$P(\rho,\Pi\rho\Pi)
\leq\frac{1}{\sqrt{\tr\rho}}\sqrt{(\tr\rho)^2-(\tr(\Pi^2\rho))^2}.$$
\end{lemma}
\begin{proof}
Note that
\begin{align*}
 \|\sqrt{\rho}\sqrt{\Pi\rho\Pi}\|_1 =\tr\sqrt{(\sqrt{\rho}\Pi\sqrt{\rho})(\sqrt{\rho}\Pi\sqrt{\rho})}=\tr(\Pi\rho) \ ,
\end{align*}
so we can write the generalized fidelity (see~\eqref{eq:genfidelity})
as
\begin{eqnarray*}
\genFid(\rho,\Pi\rho\Pi)&=&\tr(\Pi\rho)+\sqrt{(1-\tr\rho)(1-\tr(\Pi^2\rho))}\ .
\end{eqnarray*}
For brevity, we now write $\tr\rho=r$, $\tr(\Pi\rho)=s$ and
$\tr(\Pi^2\rho)=t$.  Note that $0\leq t\leq s\leq r\leq 1$. Thus,
\begin{eqnarray*}
1-\genFid(\rho,\Pi\rho\Pi)^2&=&r+t-rt-s^2-2s\sqrt{(1-r)(1-t)}.
\end{eqnarray*}
We proceed to show that $r(1-\genFid(\rho,\Pi\rho\Pi)^2)-r^2+t^2\leq 0$:
\begin{eqnarray*}
 r(1-\genFid(\rho,\Pi\rho\Pi)^2)-r^2+t^2&=&r\left(r+t-rt-s^2-2s\sqrt{(1-r)(1-t)}\right)-r^2+t^2\\
 &\leq&r\left(r+t-rt-s^2-2s(1-r)\right)-r^2+t^2\\
 &=&rt-r^2t+t^2-2rs+2r^2s-rs^2\\
 &\leq&rt-r^2t+t^2-2rs+2r^2s-rt^2\\
 &=&(1-r)(t^2+rt-2rs)\\
 &\leq&(1-r)(s^2+rs-2rs)\\
 &=&(1-r)s(s-r)\\
 &\leq&0.
\end{eqnarray*}
This completes the proof.
\end{proof}

\begin{lemma} \label{lemma:fidelity} Let $\rho\in\cU_\leq(\cH)$ and
 $\sigma\in\cU_\leq(\cH)$ have eigenvalues $r_i$ and $s_i$ ordered
 non-increasingly ($r_{i+1} \leq r_i$ and $s_{i+1} \leq s_i$). Choose
 a basis $\ket{i}$ such that $\sigma=\sum_i s_i \proj{i}$ and define
 $\tilde\rho=\sum_i r_i \proj{i}$, then
$$P(\rho, \sigma) \geq P(\tilde\rho, \sigma).$$
\end{lemma}

\begin{proof}
 By the definition of the purified distance $P(\cdot, \cdot)$, it
 suffices to show that $\genFid(\rho, \sigma) \leq \genFid(\tilde{\rho},
 \sigma)$.
\begin{align*}
 \genFid(\rho, \sigma)-\sqrt{(1-\tr\rho)(1-\tr\sigma)}& = \| \sqrt{\rho}\sqrt{\sigma}\|_1 \\
 & = \max_U \text{Re\ } \tr (U\sqrt{\rho}\sqrt{\sigma})\\
 & \leq  \max_{U, V} \text{Re\ } \tr (U\sqrt{\rho}V\sqrt{\sigma})\\
 & = \sum_i \sqrt{r_i} \sqrt{s_i} = \genFid(\tilde\rho,
 \sigma)-\sqrt{(1-\tr\tilde{\rho})(1-\tr\sigma)}.
\end{align*}
The maximizations are taken over the set of unitary matrices. The
second and third equality are Theorem~7.4.9 and Equation~(7.4.14) (on
page~436) in \cite{HornJohnsonI}.  Since $\tr\tilde{\rho}=\tr\rho$,
the result follows.
\end{proof}

\subsubsection{Basic properties of (smooth) min- and max-entropies}
\label{app:entprops}

Smooth min- and max-entropies can be seen as generalizations of the
von Neumann entropy, in the following sense~\cite{ToCoRe08s}.

\begin{lemma} \label{lem:aep} For any $\sigma\in\cU_=(\cH_{AB})$,
\begin{align*}
 \lim_{\eps \to 0} \lim_{n \to \infty} \frac{1}{n} H_{\min}^\eps(A^n|B^n)_{\sigma^{\otimes n}}
&= H(A|B)_\sigma\\
 \lim_{\eps \to 0} \lim_{n \to \infty} \frac{1}{n} H_{\max}^\eps(A^n)_{\sigma^{\otimes n}}
&=H(A)_\sigma\ .
\end{align*}
\end{lemma}

The von Neumann entropy satisfies the strong subadditivity relation,
$H(A|B C) \leq H(A | B)$. That is, discarding information encoded in a
system, $C$, can only increase the uncertainty about the state of
another system, $A$. This inequality directly generalizes to (smooth)
min- and max-entropies~\cite{RRthesiss}. In this work, we only need the
statement for $H_{\min}$.

\begin{lemma}[Strong subadditivity for
 $H_{\min}$~\cite{RRthesiss}] \label{lem:subadd} For any
 $\rho\in\cS_\leq(\cH_{ABC})$,
\begin{equation}
H_{\min}(A|B C)_{\rho|\rho}\leq H_{\min}(A|B)_{\rho|\rho}.
\end{equation}
\end{lemma}

\begin{proof}
 By definition, we have
 \begin{align*}
   2^{-H_{\min}(A|B C)_{\rho|\rho}} \id_A \ot \rho_{B C} - \rho_{A B C} \geq 0 \ .
 \end{align*}
 Because the partial trace maps nonnegative operators to nonnegative
 operators,  this implies
 \begin{align*}
    2^{-H_{\min}(A|B C)_{\rho|\rho}} \id_A \ot \rho_{B} - \rho_{A B} \geq 0 \ .
 \end{align*}
 This implies that $2^{-H_{\min}(A | B)_{\rho|\rho}} \leq 2^{-H_{\min}(A | B
     C)_{\rho|\rho}}$, which is equivalent to the assertion of the lemma.
\end{proof}

The chain rule for von Neumann entropy states that $H(A | B C) = H(A
B| C) - H(B | C)$. This equality generalizes to a family of
inequalities for (smooth) min- and max-entropies. In particular, we
will use the following two lemmas.

\begin{lemma}[Chain rule~I]
\label{lem:chain1}
For any $\rho\in\cS_\leq(\cH_{ABC})$ and $\sigma_C\in\cS_\leq(\cH_C)$,
$$H_{\min}(A|BC)_{\rho|\rho}\leq H_{\min}(AB|C)_{\rho}-H_{\min}(B|C)_{\rho} \ .$$
\end{lemma}

\begin{proof}
 Let $\sigma_C\in\cS_\leq(\cH_C)$ be arbitrary.  Then, from the
 definition of the min-entropy we have
\begin{eqnarray*}
\rho_{ABC}&\leq&2^{-H_{\min}(A|BC)_{\rho|\rho}}\id_A\ot\rho_{BC}\\
&\leq&2^{-H_{\min}(A|BC)_{\rho|\rho}}2^{-H_{\min}(B|C)_{\rho|\sigma}}\id_{AB}\ot\sigma_C.
\end{eqnarray*}
This implies that
$2^{-H_{\min}(AB|C)_{\rho|\sigma}}\leq2^{-H_{\min}(A|BC)_{\rho|\rho}}2^{-H_{\min}(B|C)_{\rho|\sigma}}$
and, hence $H_{\min}(A|B C)_{\rho|\rho}\leq H_{\min}(A
B|C)_{\rho|\sigma}-H_{\min}(B|C)_{\rho|\sigma}$. Choosing $\sigma$
such that $H_{\min}(B|C)_{\rho|\sigma}$ is maximized, we obtain
$H_{\min}(A|BC)_{\rho|\rho}\leq H_{\min}(A
B|C)_{\rho|\sigma}-H_{\min}(B|C)_{\rho}$. The desired statement then
follows because $H_{\min}(A B|C)_{\rho|\sigma} \leq H_{\min}(A
B|C)_{\rho}$.
\end{proof}

\begin{lemma}[Chain rule~II]
\label{lem:chain2}
For any $\rho\in\cS_\leq(\cH_{AB})$,
$$H_{\min}(AB)_{\rho}-\HR(B)_{\rho}\leq H_{\min}(A|B)_{\rho|\rho}.$$
\end{lemma}

Note that the inequality can be extended by conditioning all entropies
on an additional system~$C$, similarly to
Lemma~\ref{lem:chain1}. However, in this work, we only need the
version stated here.

\begin{proof}
\comment{Define $\lambda$ such that $\HR(B)=-\log_2\lambda$, $\lambda'$ such
 that $H_{\min}(A|B)=-\log_2\lambda'$, and $\lambda''$ such that
 $H_{\min}(AB)=-\log_2\lambda''$. From the definitions, we have
\begin{eqnarray*}
\rho_B\geq2^{-\HR(B)}\Pi_{\supp(\rho_B)}\\
\rho_{AB}\leq2^{-H_{\min}(AB)}\id_{AB}.
\end{eqnarray*}
and so}
From the definitions,
\begin{eqnarray*}
 \rho_{AB}&\leq&2^{-H_{\min}(AB)}\id_A\ot \Pi_{\supp(\rho_B)}\\
 &\leq&2^{-H_{\min}(AB)}2^{\HR(B)}\id_A\ot\rho_B.
\end{eqnarray*}
It follows that $2^{-H_{\min}(A|B)_{\rho|\rho}}\leq2^{-H_{\min}(AB)}2^{\HR(B)}$,
which is equivalent to the desired statement.
\end{proof}

The remaining lemmas stated in this appendix are used to transform
statements that hold for entropies $H_{\min}$ and $H_{R}$ into
statements for smooth entropies $H_{\min}^\eps$ and
$H_{\max}^\eps$. We start with an upper bound on $\HR$ in terms of
$H_{\max}$.

\begin{lemma} \label{lem:HzeroHR} For any $\eps > 0$ and for any
 $\sigma\in\cS_\leq(\cH_A)$ there exists a projector $\Pi$ which is
 diagonal in any eigenbasis of $\sigma$ such that $\tr((\id -\Pi)
 \sigma) \leq \eps$ and
\begin{align*}
 H_{\max}(A)_\sigma > \HR(A)_{\Pi \sigma \Pi} - 2 \log_2 \frac{1}{\eps} \ .
\end{align*}
\end{lemma}

\begin{proof}
 Let $\sigma = \sum_{i} r_i \proj{i}$ be a spectral decomposition of
 $\sigma$ where the eigenvalues $r_i$ are ordered non-increasingly
 ($r_{i+1} \leq r_i$).  Define the projector $\Pi_k:=\sum_{i \geq k}
 \proj{i}$. Let $j$ be the smallest index such that $\tr(\Pi_j \sigma)
 \leq \eps$ and define $\Pi := \id - \Pi_j$. Hence, $\tr(\Pi \sigma)
 \geq \tr(\sigma) - \eps$. Furthermore,
 $$\tr\sqrt{\sigma}\geq\tr(\Pi_{j-1}\sqrt{\sigma})\geq\tr(\Pi_{j-1}\sigma)\|\Pi_{j-1}\sigma\Pi_{j-1}\|_{\infty}^{-\frac{1}{2}}.$$
 We now use $\tr(\Pi_{j-1} \sigma \Pi_{j-1}) > \eps$ and the fact
 that $\|\Pi_{j-1}\sigma\Pi_{j-1}\|_{\infty}$ cannot be larger than
 the smallest non-zero eigenvalue of $\Pi \sigma \Pi$,\footnote{If
   $\Pi \sigma \Pi$ has no non-zero eigenvalue then $\HR(A)_{\Pi
     \sigma \Pi} = -\infty$ and the statement is trivial.}  which
 equals $2^{-\HR(A)_{\Pi \sigma \Pi}}$.  This implies
$$\tr\sqrt{\sigma}>\eps\sqrt{2^{\HR(A)_{\Pi \sigma \Pi}}}.$$
Taking the logarithm of the square of both sides concludes the proof.
\end{proof}

\begin{lemma} \label{lem:Hzerosmooth} For any $\eps > 0$ and for any
 $\sigma\in\cS_\leq(\cH_A)$ there exists a nonnegative operator $\Pi
 \leq \id$ which is diagonal in any eigenbasis of $\sigma$ such
 that $\tr((\id-\Pi^2) \sigma) \leq 2 \eps$ and
\begin{align*}
 H_{\max}^\eps(A)_{\sigma} \geq H_{\max}(A)_{\Pi \sigma \Pi} \ .
\end{align*}
\end{lemma}

\begin{proof} By definition of $H_{\max}^\eps(A)_{\sigma}$, there is a
 $\rho\in\cB^\eps(\sigma)$ such that $H_{\max}^\eps(A)_{\sigma}=
 H_{\max}(A)_{\rho}$.  It follows from Lemma~\ref{lemma:fidelity}
 that we can take $\rho$ to be diagonal in any eigenbasis of
 $\sigma$. Define
 \begin{align*}
   \rho':=\rho-\{\rho-\sigma\}_{+}=\sigma-\{\sigma-\rho\}_{+}
 \end{align*}
 where $\{\cdot\}_{+}$ denotes the positive part of an operator. We
 then have $\rho'\leq\rho$, which immediately implies that
 $H_{\max}(A)_{\rho'} \leq H_{\max}(A)_{\rho}$. Furthermore, because
 $\rho'\leq\sigma$ and because $\rho'$ and $\sigma$ have the same
 eigenbasis, there exists a nonnegative operator $\Pi\leq\id$
 diagonal in the eigenbasis of $\sigma$ such that
 $\rho'=\Pi\sigma\Pi$. The assertion then follows because
 \begin{align*}
   \tr((\id - \Pi^2)\sigma)=\tr(\sigma)-\tr(\rho')
 =\tr(\{\sigma-\rho\}_{+})
 \leq\|\rho-\sigma\|_1
 \leq 2\eps \ ,
 \end{align*}
 where the last inequality follows from Lemma~\ref{lem:Ptracebound}
 and $P(\rho,\sigma) \leq \eps$.
\end{proof}


\begin{lemma} \label{lem:smoothHR} For any $\eps > 0$ and for any
 $\sigma\in\cS_\leq(\cH_A)$ there exists a nonnegative operator $\Pi
 \leq \id$ which is diagonal in any eigenbasis of $\sigma$ such that
 $\tr((\id - \Pi^2) \sigma) \leq 3 \eps$ and
\begin{align*}
 H_{\max}^\eps(A)_{\sigma} \geq \HR(A)_{\Pi \sigma \Pi} - 2 \log_2
 \frac{1}{\eps} \ .
\end{align*}
\end{lemma}

\begin{proof}
 By Lemma~\ref{lem:Hzerosmooth}, there exists a nonnegative operator
 $\bar{\Pi} \leq \id$ such that
 $$H_{\max}^\eps(A)_{\sigma} \geq H_{\max}(A)_{\bar{\Pi}\sigma \bar{\Pi}}$$
 and $\tr((\id - \bar{\Pi}^2)\sigma) \leq 2 \eps$. By
 Lemma~\ref{lem:HzeroHR} applied to $\bar{\Pi}\sigma\bar{\Pi}$,
 there exists a projector $\bar{\bar{\Pi}}$ such that
 $$H_{\max}(A)_{\bar{\Pi} \sigma \bar{\Pi}} \geq \HR(A)_{\Pi \sigma \Pi} -
 2 \log_2 \frac{1}{\eps}$$ and $\tr((\id - \bar{\bar{\Pi}} )
 \bar{\Pi} \sigma \bar{\Pi}) \leq \eps$, where we defined $\Pi :=
 \bar{\bar{\Pi}} \bar{\Pi}$. Furthermore, $\bar{\Pi}$,
 $\bar{\bar{\Pi}}$ and, hence, $\Pi$, can be chosen to be diagonal in
 any eigenbasis of $\sigma$.  The claim then follows because
 $$\tr((\id - \Pi^2)\sigma) = \tr((\id - \bar{\bar{\Pi}} \bar{\Pi}^2) \sigma) = \tr((\id - \bar{\Pi}^2) \sigma) + \tr((\id - \bar{\bar{\Pi}} ) \bar{\Pi} \sigma\bar{\Pi}) \leq 3 \eps.$$
\end{proof}

\begin{lemma} \label{lem:Hmaxmonotonicity} Let $\eps \geq 0$, let
 $\sigma\in\cS_\leq(\cH_A)$ and let $\cM : \, \sigma \mapsto \sum_{i}
 \proj{\phi_i} \bra{\phi_i} \sigma \ket{\phi_i}$ be a measurement
 with respect to an orthonormal basis $\{\ket{\phi_i}\}_{i}$. Then
 \begin{align*}
   H_{\max}^\eps(A)_{\sigma} \leq H_{\max}^\eps(A)_{\cM(\sigma)} \ .
 \end{align*}
\end{lemma}

\begin{proof}
 The max-entropy can be written in terms of the (standard) fidelity
 (see also~\cite{KoReSc08}) as $$H_{\max}(A)_{\sigma}=2 \log_2
 F(\sigma_A, \id_A).$$ Using the fact that the fidelity can only
 increase when applying a trace-preserving completely positive map
 (see, e.g., \cite{NieChu}), we have
 \begin{align*}
   F(\sigma_A, \id_A) \leq F(\cM(\sigma_A), \cM(\id_A)) = F(\cM(\sigma_A), \id_A) \ .
 \end{align*}
 Combining this with the above yields
 \begin{align} \label{eq:nonsmoothHmaxdecrease}
   H_{\max}(A)_{\sigma} \leq H_{\max}(A)_{\cM(\sigma)} \ ,
 \end{align}
 which proves the claim in the special case where $\eps = 0$.

 To prove the general claim, let $\cH_{\obst}$ and $\cH_{\obst'}$ be
 isomorphic to $\cH_A$ and let $U$ be the isometry from $\cH_A$ to
 $\mathrm{span} \{\ket{\phi_i}_\obst \otimes \ket{\phi_i}_{\obst'}\}_i
 \subseteq \cH_\obst \otimes \cH_{\obst'}$ defined by $\ket{\phi_i}_A \to
 \ket{\phi_i}_{\obst} \otimes \ket{\phi_i}_{\obst'}$. The action of $\cM$ can
 then equivalently be seen as that of $U$ followed by the partial
 trace over $\cH_{\obst'}$. In particular, defining $\sigma'_{\obst \obst'} := U
 \sigma_A U^{\dagger}$, we have $\cM(\sigma_A) = \sigma'_\obst$.

 Let $\rho'\in\cS(\cH_{\obst\obst'})$ be a density operator such that
 \begin{align} \label{eq:Hmaxsmootheq}
   H_{\max}(\obst)_{\rho'}=H_{\max}^\eps(\obst)_{\sigma'}
 \end{align}
 and
 \begin{align} \label{eq:distXX}
   P(\rho'_{\obst\obst'}, \sigma'_{\obst\obst'}) \leq \eps \ .
 \end{align}
 (Note that, by definition, there exists a state $\rho'_\obst$ that
 satisfies~\eqref{eq:Hmaxsmootheq} with $P(\rho'_\obst, \sigma'_\obst) \leq
 \eps$.  It follows from Uhlmann's theorem (see e.g.~\cite{NieChu})
 and the fact that the purified distance is non-increasing under
 partial trace that there exists an extension of $\rho'_\obst$ such
 that~\eqref{eq:distXX} also holds.)

 Since $\sigma'_{\obst \obst'}$ has support in the subspace $\mathrm{span}
 \{\ket{\phi_i}_\obst \otimes \ket{\phi_i}_{\obst'}\}_i$, we can assume that
 the same is true for $\rho'_{\obst \obst'}$.  To see this, define $\Pi$ as
 the projector onto this subspace and observe that
 $\tr_{\obst'}(\Pi\rho'_{\obst\obst'}\Pi)$ cannot be a worse candidate for the
 optimization in $H_{\max}^{\eps}(\obst)_{\sigma'}$: From
 Lemma~\ref{lemma:fidelity}, we can take $\rho'_\obst$ to be diagonal in
 the $\{\ket{\phi_i}\}$ basis, i.e.\ we can write
 $$\rho'_\obst=\sum_i\lambda_i\ketbra{\phi_i}{\phi_i},$$ where
 $\lambda_i\geq 0$.  We also
 write $$\rho_{\obst\obst'}=\sum_{ijkl}c_{ijkl}\ketbra{\phi_i}{\phi_j}\ot\ketbra{\phi_k}{\phi_l},$$
 for some coefficients $c_{ijkl}$.  To ensure
 $\rho'_\obst=\tr_{\obst'}\rho'_{\obst\obst'}$, we require
 $\sum_kc_{ijkk}=\lambda_i\delta_{ij}$.  Consider then
 \begin{eqnarray*}
   \tr_{\obst'}(\Pi\rho'_{\obst\obst'}\Pi)&=&\tr_{\obst'}\left(\sum_{ij}c_{ijij}\ketbra{\phi_i}{\phi_j}\ot\ketbra{\phi_i}{\phi_j}\right)\\
   &=&\sum_ic_{iiii}\ketbra{\phi_i}{\phi_i}.
 \end{eqnarray*}
 It follows that $\tr_{\obst'}(\Pi\rho'_{\obst \obst'}\Pi)\leq\rho'_\obst$ (since
 $\sum_kc_{iikk}=\lambda_i$ and $c_{iikk}\geq 0$) and hence we
 have $$H_{\max}(\obst)_{\tr_{\obst'}(\Pi\rho'_{\obst \obst'}\Pi)}\leq
 H_{\max}^{\eps}(\obst)_{\rho'}.$$ Furthermore, from
 Lemma~\ref{lem:Pnonincreasing}, we have
 $$P(\Pi\rho'_{\obst \obst'}\Pi,\sigma'_{\obst\obst'})=P(\Pi\rho'_{\obst
   \obst'}\Pi,\Pi\sigma'_{\obst \obst'}\Pi)\leq P(\rho'_{\obst \obst'}, \sigma'_{\obst
   \obst'})\leq\eps,$$ from which it follows
 that $$\tr_{\obst'}(\Pi\rho'_{\obst\obst'}\Pi)\in\cB^{\eps}(\sigma'_\obst).$$ We
 have hence shown that there exists a state $\rho'_{\obst\obst'}$
 satisfying~\eqref{eq:Hmaxsmootheq} and~\eqref{eq:distXX} whose
 support is in $\mathrm{span}\{\ket{\phi_i}_\obst \otimes
 \ket{\phi_i}_{\obst'}\}_i$.

 We can thus define $\rho_A := U^{\dagger} \rho'_{\obst \obst'} U$ so that
 $\rho'_\obst = \cM(\rho_A)$ and hence~\eqref{eq:Hmaxsmootheq} can be
 rewritten as
 \begin{align*}
   H_{\max}(A)_{\cM(\rho)} = H_{\max}^\eps(A)_{\cM(\sigma)} \ ,
 \end{align*}
 and~\eqref{eq:distXX} as
 \begin{align*}
   P(\rho_A, \sigma_A) \leq \eps \ .
 \end{align*}
 Using this and~\eqref{eq:nonsmoothHmaxdecrease}, we conclude that
 \begin{align*}
   H_{\max}^\eps(A)_{\cM(\sigma)}=H_{\max}(A)_{\cM(\rho)}\geq
   H_{\max}(A)_{\rho}\geq H_{\max}^\eps(A)_{\sigma} \ .
 \end{align*}
\end{proof}

\begin{lemma} \label{lem:Hzerosmoothsubstate} Let $\eps \geq 0$, and
 let $\sigma\in\cS_\leq(\cH_A)$ and $\sigma'\in\cS_\leq(\cH_A)$. If
 $\sigma'\leq \sigma$ then
 \begin{align*}
   H_{\max}^{\eps}(A)_{\sigma'} \leq H_{\max}^\eps(A)_{\sigma}\ .
 \end{align*}
\end{lemma}

\begin{proof}
 By Lemma~\ref{lem:Hmaxmonotonicity}, applied to an orthonormal
 measurement $\cM$ with respect to the eigenbasis of $\sigma$, we
 have
 \begin{align*}
   H_{\max}^\eps(A)_{\sigma'}
 \leq
   H_{\max}^\eps(A)_{\cM(\sigma')} \ .
 \end{align*}
 Using this and the fact that $\cM(\sigma') \leq \cM(\sigma) =
 \sigma$, we conclude that it suffices to prove the claim for the
 case where $\sigma'$ and $\sigma$ are diagonal in the same
 basis.

 By definition, there exists $\rho$ such that $P(\rho, \sigma) \leq
 \eps$ and $H_{\max}(A)_{\rho} = H_{\max}^\eps(A)_{\sigma}$. Because
 of Lemma~\ref{lemma:fidelity}, $\rho$ can be assumed to be diagonal
 in an eigenbasis of $\sigma$. Hence, there exists an operator
 $\Gamma$ which is diagonal in the same eigenbasis such that $\rho =
 \Gamma \sigma \Gamma$.  We define $\rho' := \Gamma \sigma' \Gamma$
 for which $\rho' \geq 0$ and $\tr(\rho') \leq \tr(\rho) \leq
 1$. Furthermore, since $\rho' \leq \rho$, we have
  \begin{align*}
    H_{\max}(A)_{\rho'} \leq H_{\max}(A)_{\rho} = H_{\max}^\eps(A)_{\sigma} \ .
  \end{align*}
  Because $\sigma'$ and $\sigma$ can be assumed to be diagonal in
  the same basis, there exists a nonnegative operator $\Pi \leq \id$
  which is diagonal in the eigenbasis of $\sigma$ (and, hence, of
  $\Gamma$ and $\rho$) such that $\sigma' = \Pi \sigma \Pi$. We
  then have
  \begin{align*}
    \rho' = \Gamma \sigma' \Gamma = \Gamma \Pi \sigma \Pi \Gamma = \Pi \Gamma \sigma \Gamma \Pi = \Pi \rho \Pi \ .
  \end{align*}
  Using the fact that the purified distance can only decrease under
  the action of $\Pi$ (see Lemma~\ref{lem:Pnonincreasing}), we have
  \begin{align*}
    P(\rho', \sigma') = P(\Pi \rho \Pi, \Pi \sigma \Pi) \leq P(\rho,
  \sigma) \leq \eps \ .
  \end{align*}
  This implies $H_{\max}^\eps(A)_{\sigma'} \leq H_{\max}(A)_{\rho'}$
  and thus concludes the proof.
\end{proof}

\begin{lemma} \label{lem:smoothHmin} For any $\eps \geq 0$ and for any
 (normalized) $\sigma\in\cS_=(\cH_A)$, there exists a nonnegative
 operator $\Pi \leq \id$ which is diagonal in any eigenbasis of
 $\sigma$ such that $\tr((\id-\Pi^2)\sigma) \leq 2 \eps$ and
\begin{align*}
 H_{\min}^\eps(A)_{\sigma}\leq H_{\min}(A)_{\Pi \sigma \Pi}\ .
\end{align*}
\end{lemma}

\begin{proof}
 Let $\rho\in \cB^\eps(\sigma)$ be such that
 $H_{\min}(A)_{\rho}=H^{\eps}_{\min}(A)_{\sigma}$.  It follows from
 Lemma~\ref{lemma:fidelity} that we can take $\rho$ to be diagonal
 in an eigenbasis $\ket{i}$ of $\sigma$. Let $r_i$ ($s_i$) be the
 list of eigenvalues of $\rho$ ($\sigma$) and define
 $\sigma'_A=\sum_i \min(r_i, s_i) \proj{i}$. It is easy to see that
 there exists a nonnegative operator $\Pi\leq\id$ such that
 $\sigma'= \Pi \sigma \Pi$. Since $\sigma' \leq \rho$, we
 have
 \begin{align*}
   H_{\min}(A)_{\Pi \sigma \Pi}=H_{\min}(A)_{\sigma'}\geq
   H_{\min}(A)_{\rho}=H^{\eps}_{\min}(A)_{\sigma}\ .
 \end{align*}
 Furthermore, $\tr((\id - \Pi^2)\sigma) = \tr(\sigma - \sigma')
 = \sum_{i: \, s_i \geq r_i} (s_i - r_i) \leq \| \sigma - \rho
 \|_1$. The assertion then follows because, by
 Lemma~\ref{lem:Ptracebound}, the term on the right hand side is
 bounded by $2 P(\sigma, \rho) \leq 2 \eps$.
\end{proof}



\end{widetext}

\end{document}